\documentstyle[aps,pre,preprint,tighten,epsf,epsfig]{revtex}

\newcommand{\cn}{{\rm cn}}

\newcommand{\cd}{{\rm cd}}

\newcommand{\Sc}{{\rm sc}}
\newcommand{\F}{{\rm F}}
\newcommand{\K}{{\rm K}}
\newcommand{\E}{{\rm E}}

\newcommand{\cA}{{\cal A}}
\newcommand{\cV}{{\cal V}}
\newcommand{\xc}{x_{\rm c}}
\newcommand{\qt}{q_{\rm t}}
\newcommand{\qc}{q_{\rm c}}
\newcommand{\tqt}{\tilde{q}_{\rm t}}
\newcommand{\xgm}{x_{\rm gm}}
\newcommand{\xlm}{x_{\rm lm}}
\newcommand{\xsp}{x_{\rm sp}}
\newcommand{\xct}{x_{\rm ct}}

\begin{document}

\draft

\title{Improved semiclassical density matrix:
taming caustics}
\author{C.\ A.\ A.\ de Carvalho,$^{1,}$\footnote{E-mail:
aragao@if.ufrj.br}
R.\ M.\ Cavalcanti,$^{1,2,}$\footnote{E-mail: rmoritz@if.ufrj.br}
E.\ S.\ Fraga,$^{3,}$\footnote{Present address:
Laboratoire de Physique Th\'eorique,
Universit\'e Paris-Sud XI,
B\^atiment 210, 91405 Orsay Cedex, France;
E-mail: fraga@th.p-sud.fr} and
S.\ E.\ Jor\'as$^{4,}$\footnote{Present address: Instituto de F\'{\i}sica,
Universidade Fe\-de\-ral do Rio de Janeiro,
Caixa Postal 68528, Rio de Janeiro, RJ 21945-970, Brazil;
E-mail: joras@if.ufrj.br}}
\address{$^{1}$Instituto de F\'\i sica,
Universidade Federal do Rio de Janeiro, \\
Caixa Postal 68528, Rio de Janeiro, RJ 21945-970, Brazil \\
$^{2}$Instituto de F\'\i sica, Universidade de S\~ao Paulo, \\
Caixa Postal 66318, S\~ao Paulo, SP 05315-970, Brazil \\
$^{3}$Nuclear Theory Group, Physics Department,
Brookhaven National Laboratory, \\
Upton, NY 11973-5000, U.S.A. \\
$^{4}$High Energy Theory Group, Physics Department,
Brown University, \\
Providence, RI 02912, U.S.A.}
\maketitle

\begin{abstract}

We present a simple method
to deal with caustics in the semiclassical
approximation to the thermal density matrix of a
particle moving on the line.
For simplicity, only its diagonal elements are
considered.
The only ingredient we require is the knowledge
of the extrema of the Euclidean
action. The procedure makes use of
complex trajectories, and is applied to the quartic double-well
potential.

\end{abstract}



\section{Introduction}

In the path integral formulation of quantum statistical
mechanics, the thermal density matrix
$\rho(x,x')=\langle x|\exp(-\beta H)|x'\rangle$
of a system with Hamiltonian
\begin{equation}
H=\frac{p^2}{2m}+V(x)
\end{equation}
is given by \cite{Feynman,Schulman,Kleinert}
\begin{equation}
\label{rho1}
\rho(x,x')=\int_{z(0)=x'}^{z(\beta\hbar)=x}
[{\cal D}z(\tau)]\,\exp\left(-\frac{S[z]}{\hbar}\right),
\end{equation}
with
\begin{equation}
S[z]=\int_0^{\beta\hbar} d\tau
\left[\frac{1}{2}\,m\dot{z}^2+V(z)\right].
\end{equation}
A semiclassical series for $\rho(x,x')$ may be obtained from Eq.\
(\ref{rho1}) through the method of steepest descent. The
derivation depends solely on the knowledge of the paths that are
{\em minima} of the Euclidean action $S$ (the Euclidean nature of
the path integral allows us to discard saddle-points). They act
as backgrounds upon which a semiclassical propagator can be obtained
exactly and then used to construct the series perturbatively.
Its first term is given by
\begin{equation}
\label{sc}
\rho_{\Sc}^{(1)}(x,x')=
\sum_{j=1}^{N}\exp\left(-\frac{S[\xc^j]}{\hbar}\right)
\Delta_j^{-1/2}.
\end{equation}
The sum runs over all minima $\xc^j(\tau)$ of the action $S[z]$
satisfying the boundary conditions $z(0)=x'$ and
$z(\beta\hbar)=x$, and $\Delta_j$ denotes the determinant of
\begin{equation}
\label{F}
\hat{F}[\xc^j]\equiv -m\,\frac{d^2}{d\tau^2}+V''[\xc^j],
\end{equation}
the operator of quadratic fluctuations around $\xc^j(\tau)$.
(A derivation of this result will be sketched in Sec.\ \ref{improved}.)

In previous works \cite{CAAC2,CAAC3},
we presented the explicit construction of the
series for the diagonal elements of the density matrix,
$\rho(x,x)$. For the sake of simplicity, we
restricted our discussion to potentials of the single-well type.
The more intricate case of multiple-wells --- of which the quartic
double well, with its many applications of practical importance
\cite{Miller}, is a paradigm --- was left aside, as it requires
special treatment. Differently from single-wells, for
multiple-wells the number $N$ of minima of $S$ depends on
$x$ and $\beta$ \cite{CAAC1}. On the frontier separating regions in
the $(x,\beta)$-plane with different values of $N$ ---
a {\em caustic} --- $\rho_{\Sc}(x,x)$ diverges due to the vanishing of
the fluctuation determinant around the minimum that appears
or disappears there. In the present
context, this divergence is an artifact of
the semiclassical approximation.
Thus, a simple manner of eliminating it is certainly called for;
this is the purpose of this paper.

As the caustic problem appears in other contexts in physics,
it is instructive to briefly review how it comes about, and how
it has been dealt with, for the sake of comparison.
In optics, caustics occur whenever light rays coalesce.
Thus, they separate regions of different number of extrema
(the light rays) of the optical distance (the analogue of the
action). In order to go beyond geometrical optics, one
has to take into account fluctuations around these light rays.
Just as in the present case, singularities emerge when we
compute quadratic fluctuations on caustics.
Ways to avoid this have been known for some
time \cite{Berry3,Trinkaus,Berry2,Berry4,Berry5}. Indeed, due
to the traditional analogy between wave optics and quantum
mechanics, the techniques involved are similar to the ones
used in deriving connection formulae for WKB
approximations \cite{balianbloch}, and consist essentially in
replacing one or more of the Fresnel integrals that arise in
the stationary phase approximation with a so-called
diffraction integral, whose form is specified by the
classification of the caustic according to
Catastrophe Theory \cite{Thom,Poston}; in the simplest
case it is an Airy-type integral.
A general procedure has also been developed to
deal with caustics in the path integral formulation
of quantum mechanics \cite{DV}. Although
this general procedure could, in principle, be adapted to
the case at hand, the nature of our problem allows for
simplifications which warrant special treatment.

In {\em nonequilibrium} quantum statistical mechanics,
caustics are also known to occur in semiclassical descriptions
of the decay of metastable states.
The problem here is that of a particle in a potential which
has a local minimum separated by a barrier from a region
where it is unbounded below, in contact with a thermal
reservoir. The phenomenon of caustics
has been associated with a transition from the
classical to the quantum regime of the decay
rate \cite{Affleck,Chudnovsky,Weiss}.
General prescriptions
for dealing with this phenomenon near the top of the
barrier have been given in great
detail \cite{Weiss,Ankerhold,Weiper1}.

The case we shall analyze in this article differs from the
one in the previous paragraph in the following aspects:
(i) we discuss a problem in {\em equilibrium} quantum
statistical mechanics; (ii) our analysis
is global, in the sense that we compute the density matrix
diagonal for every point on the real axis.
It also differs from analyses carried
out in optics and quantum mechanics because of the
Euclidean nature of the path integral: only minima are
to be considered; saddle-points are discarded. (This is strictly
true only in the ``usual'' semiclassical approximation.
Our ``improved'' approximation makes use of some of the
saddle-points.)
In fact, in the specific example
we analyze (the quartic double-well potential) only one new (local)
minimum is introduced in function space after the various
catastrophes that occur as we change the temperature;
as it appears after the first catastrophe, this is the
only one we have to consider. (Again, this is valid only
for the ``usual'' semiclassical approximation; the improved one
also requires an analysis of the second catastrophe.)
This results in a
prescription for dealing with caustics that is
simple and direct.

Previous works have studied
the quartic double-well potential at finite temperature using
semiclassical methods \cite{Harrington,Dolan}. Our work
complements and extends those studies, by giving an explicit
recipe for dealing with caustics. Variational methods
have also proven extremely useful in this problem, and
were quite successful in addressing applications in
condensed matter physics \cite{Tognetti}. Combinations of
perturbation theory with variational techniques have
also been recently used \cite{Bachmann}. Our contribution
to the semiclassical treatment opens the way for
practical calculations, to be compared with
perturbative, variational and numerical results.

This article is organized as follows: before introducing
the improved
semiclassical approximation, which will remedy the problem
of spurious divergences, Sec.\ \ref{improved} briefly
reviews the derivation of the usual
semiclassical approximation to the density matrix.
The new method
is, then, presented in two alternative ways: one has a better
physical motivation, and clearly illustrates the essential
ideas, but lacks effectiveness as a calculational tool; the
other provides a general recipe to perform the calculations
in a systematic way, resorting to the use of complex trajectories.
Sec.\ \ref{application} describes in detail the results obtained
for the case of
the quartic double-well potential by using the improved semiclassical
approximation, which are compared to the usual approach.
Sec.\ \ref{conclusion}
presents our conclusions and points out directions for future work.


\section{Improving the semiclassical approximation}
\label{improved}

\subsection{The ``usual'' semiclassical approximation}

In order to show how one can improve the semiclassical
approximation so as to eliminate the unphysical
divergences at the caustics, it is convenient to
remember how the usual semiclassical approximation
to a path integral like the one in Eq.\ (\ref{rho1})
is derived. Briefly, one has to:

(A) Solve the Euler-Lagrange equation, $m\ddot{z}=V'(z)$,
subject to the boundary conditions
$z(\beta\hbar)=z(0)=x$, and determine, among the solutions,
those which {\em minimize} (globally or locally) the action.
For simplicity, we shall assume for the moment that there
is only one such solution, which we denote by $\xc$;

(B) Expand the action around $\xc$:
$S[\xc+\eta]=S[\xc]+S_2+\delta S$, where
\begin{equation}
S_2 = \frac{1}{2}\int_0^{\beta\hbar}d\tau\,\eta(\tau)
\,\hat{F}[\xc]\,\eta(\tau),
\label{S_2}
\end{equation}
\begin{equation}
\delta S = \sum_{k=3}^{\infty}\frac{1}{k!}\int_0^{\beta\hbar}
d\tau\,V^{(k)}[\xc(\tau)]\,\eta^k(\tau);
\end{equation}
$\hat{F}$ is the operator defined in Eq.\ (\ref{F}), and
we are assuming that $V(z)$ is an analytic function of $z$,
so that all derivatives $V^{(k)}(z)$ exist;

(C) Expand the fluctuations $\eta(\tau)$ in terms of the orthonormal
modes of the fluctuation operator $\hat{F}[\xc]$:
\begin{equation}
\eta(\tau)=\sum_{j=0}^{\infty}a_j\varphi_j(\tau),
\end{equation}
where $\hat{F}\varphi_j(\tau)=\lambda_j\varphi_j(\tau)$, with
$\varphi_j(0)=\varphi_j(\beta\hbar)=0$; then
\begin{equation}
S_2 = \frac{1}{2}\sum_{j=0}^{\infty}\lambda_ja_j^2,
\end{equation}
\begin{equation}
\label{deltaS}
\delta S = \sum_{n=3}^{\infty}\sum_{i_1=0}^{\infty}\cdots
\sum_{i_n=0}^{\infty}
\frac{1}{n!}\,C^{(n)}_{i_1i_2\ldots i_n}\,a_{i_1}a_{i_2}\ldots a_{i_n},
\end{equation}
where
\begin{equation}
\label{C}
C^{(n)}_{i_1\ldots i_n}=\int_0^{\beta\hbar}d\tau\,
V^{(n)}[\xc(\tau)]\,\varphi_{i_1}(\tau)
\ldots\varphi_{i_n}(\tau).
\end{equation}

The ``usual'' semiclassical approximation
is obtained by neglecting $\delta S$ in the path integral
on the r.h.s.\ of Eq.\ (\ref{rho1}), which, upon the
change of variables $\eta(\tau)\to\{a_j\}$, becomes a
product of Gaussian integrals:
\begin{equation}
\rho(x,x)\approx \exp\left(-\frac{S[\xc]}{\hbar}\right)
\prod_{j=0}^{\infty}{\cal I}_j\:,
\end{equation}
\begin{equation}
{\cal I}_j=\int_{-\infty}^{\infty}\frac{da_j}{\sqrt{2\pi\hbar}}\,
\exp\left(-\frac{\lambda_j^{}a_j^2}{2\hbar}\right)=\lambda_j^{-1/2}.
\end{equation}
Hence
\begin{equation}
\label{exp1}
\rho(x,x)\approx\exp\left(-\frac{S[\xc]}{\hbar}\right)\Delta^{-1/2},
\end{equation}
where $\Delta=\prod_{j=0}^{\infty}\lambda_j={\rm det}\,\hat{F}$.
(Explicit expressions for $\Delta$ [see Eq.\ (\ref{Delta}) below]
were derived in Ref.\ \cite{CAAC2}, where it was also discussed how to
systematically include corrections due to $\delta S$.)
If there are $N$ minima, one has to add together their
contributions, thus obtaining Eq.\ (\ref{sc}).


\subsection{Taming the caustics}

When we cross a caustic, a classical trajectory $\xc(\tau)$
is created or annihilated. Precisely at this point, the
lowest eigenvalue of $\hat{F}[\xc]$ vanishes, thus making
the integral ${\cal I}_0$
blow up. This problem can be remedied by retaining
fluctuations beyond quadratic in the subspace spanned
by $\varphi_0$ (the eigenmode of $\hat{F}$ associated
with $\lambda_0$), i.e., we replace ${\cal I}_0$ with
\begin{equation}
\label{calF}
\widetilde{\cal I}_0=\int_{-\infty}^{\infty}\frac{da_0}{\sqrt{2\pi\hbar}}
\,\exp\left(-\frac{\cV(a_0)}{\hbar}\right)
\equiv\lambda_0^{-1/2}\,{\cal F},
\end{equation}
where
\begin{equation}
\label{U}
\cV(a_0)=\frac{1}{2}\,\lambda_0a_0^2+\sum_{n=3}^{M}\frac{1}{n!}\,
C^{(n)}_{00\ldots 0}\,a_0^n.
\end{equation}
We take for $M$ the smallest {\em even} integer such that
$C^{(M)}_{00\ldots 0}$
is positive for all values of $x_0$ and $\beta$; this suffices
to make the integral in (\ref{calF}) finite
even when $\lambda_0$ vanishes.

As a result, we obtain an improved approximation to
the density matrix element (\ref{rho1}):
\begin{equation}
\label{exp2}
\rho(x,x)
\approx \exp\left(-\frac{S[\xgm]}{\hbar}\right)
\Delta^{-1/2}\,{\cal F}.
\end{equation}
Here, $\xgm$ is the {\it global minimum} of $S[x]$.

It is important to note that there is a one-to-one
correspondence between the minima of $S[x]$ and the
minima of $\cV(a_0)$. Therefore, it is not necessary
to explicitly add their contributions as in Eq.\ (\ref{sc}),
for they are already included in (\ref{exp2}).
Indeed, let $a_{01}(=0), a_{02}, \ldots, a_{0N}$ be the minima of
$\cV$. If they are sufficiently far apart, one
may compute ${\cal F}$ using the steepest descent
method, obtaining
\begin{equation}
{\cal F}\sim \sum_{j=1}^{N}\sqrt{\frac{\lambda_0}
{\cV''(a_{0j})}}\,\exp\left(-\frac{\cV(a_{0j})}{\hbar}\right)
\end{equation}
Substituting this into Eq.\ (\ref{exp2}) then yields
\begin{equation}
\label{Fasymp}
\rho(x,x)
\approx \sum_{j=1}^{N}\exp\left(-\frac{S_j}{\hbar}\right)
\widetilde{\Delta}_j^{-1/2},
\end{equation}
where $S_j\equiv S[\xgm]+\cV(a_{0j})$ and
$\widetilde{\Delta}_j\equiv(\cV''(a_{0j})/\lambda_0)\,\Delta$.
Although expression (\ref{Fasymp}) is not identical to the
``usual'' semiclassical approximation [Eq.\ (\ref{sc})],
the dominant term in both sums is the same, namely
$\exp(-S[\xgm]/\hbar)\,\Delta^{-1/2}$
(recall that $a_{01}=0$, hence $\cV(a_{01})=0$ and
$\cV''(a_{01})=\lambda_0$); the other terms are
exponentially supressed in the classical limit $\hbar\to 0$.

Another point that is important to mention is that one can,
in principle, systematically improve the ``improved'' semiclassical
approximation, Eq.\ (\ref{exp2}). To do this, one first
decomposes the action into three pieces: $S[\xgm+\eta]=S[\xgm]+
S_{\rm I}+S_{\rm II}$, where
\begin{equation}
S_{\rm I}=\cV(a_0)+\frac{1}{2}\sum_{j=1}^{\infty}\lambda_ja_j^2
\end{equation}
and
\begin{equation}
S_{\rm II}=\delta S-\left[\cV(a_0)-\frac{1}{2}\,
\lambda_0a_0^2\right],
\end{equation}
with $\delta S$ defined by Eq.\ (\ref{deltaS}).
Applying this decomposition to Eq.\ (\ref{rho1}) (with $x'=x$)
then yields
\begin{equation}
\label{impseries}
\rho(x,x)=\exp\left(-\frac{S[\xgm]}{\hbar}\right)\int\prod_{j=0}^{\infty}
\frac{da_j}{\sqrt{2\pi\hbar}}\,\exp\left(-\frac{S_{\rm I}}{\hbar}\right)
\sum_{n=0}^{\infty}\frac{1}{n!}
\left(-\frac{S_{\rm II}}{\hbar}\right)^n.
\end{equation}
This defines an ``improved semiclassical series,''
the first term of which corresponds to Eq.\ (\ref{exp2}).
Higher order terms can be readily computed,
as they can be recast as sums of products of
simple integrals. Compared with the ``usual'' semiclassical
series \cite{CAAC2}, the series (\ref{impseries})
has the disadvantage that
integrals involving powers of $a_0$ must be computed
numerically. On the other hand, those integrals are
finite even at the caustics, so that the coefficients of
the series (\ref{impseries}) are well defined for
any $\beta$ and $x_0$.

Although the procedure outlined in this subsection
teaches us how to
cross the caustics, it is not very convenient:
in order to obtain the coefficients of $\cV(a_0)$
one has to find $\lambda_0$ and $\varphi_0(\tau)$.
This, in general, is not an easy
task, and makes the whole procedure very cumbersome.
Instead, we shall present an
alternative way of obtaining those coefficients,
which is based on the one-to-one correspondence
between the minima of $S$ and $\cV$.


\subsection{An alternative procedure}
\label{alternative}

Let us assume that $M=4$ in Eq.\ (\ref{U}); this is the case
for the quartic double-well potential, to be discussed in
the next section. Then the ``effective action''
$\cA(a_0)\equiv S[\xgm]+\cV(a_0)$
for the ``critical'' mode $\varphi_0$ is a
fourth degree polynomial in $a_0$.

Let us also assume for the moment that $\cA(a_0)$
has three extrema:
a global minimum at $a_0=0$, a local maximum
at $u>0$, and a local minimum at $v>u$.
This allows us to write $\cA(a_0)$ as
\begin{equation}
\label{S2}
\cA(a_0)=S[\xgm]+\alpha\left[\frac{1}{2}\,uv\,a_0^2
-\frac{1}{3}\,(u+v)\,a_0^3+\frac{1}{4}\,a_0^4\right]
\end{equation}
(one can easily check that $\cA'(0)=\cA'(u)=\cA'(v)=0$).

We now have to relate $\alpha$, $u$, and $v$ to
calculable quantities. We do this by imposing that
$\cA(v)=S[\xlm]$ and $\cA(u)=S[\xsp]$, where $\xlm(\tau)$
and $\xsp(\tau)$ are the local minimum and the lowest
saddle-point of $S[x]$, respectively.
This yields
\begin{equation}
\label{xi}
\frac{S[\xlm]-S[\xgm]}{S[\xsp]-S[\xgm]}
=\frac{\cA(v)-\cA(0)}{\cA(u)-\cA(0)}
=\frac{\xi^3(2-\xi)}{2\xi-1},
\end{equation}
where $\xi\equiv v/u$.
It follows from the definition of $\xgm$, $\xlm$ and $\xsp$ that
the l.h.s.\ of Eq.\ (\ref{xi}) is in the range $[0,1]$. A plot of its
r.h.s.\ shows that Eq.\ (\ref{xi}) possesses a unique real solution,
lying in the interval $1\le\xi\le 2$.

Having determined $\xi$, we can now fix another combination
of parameters, namely $\mu\equiv\alpha u^4$:
\begin{equation}
\label{lambda}
S[\xsp]-S[\xgm]=\cA(u)-\cA(0)=\frac{\mu}{12}\,(2\xi-1).
\end{equation}
We can then rewrite (\ref{S2}) as
$\cA(a_0)=S[\xgm]+\cV_3(a_0/u)$, where
\begin{equation}
\label{U3}
\cV_3(z)\equiv\mu\left[\frac{1}{2}\,\xi z^2
-\frac{1}{3}\,(1+\xi)\,z^3+\frac{1}{4}\,z^4\right].
\end{equation}
There still remains one parameter to be determined,
namely $u$. Fortunately, we do not need it in order
to compute ${\cal F}$. Indeed,
identifying $\cV_3(a_0/u)$ with $\cV(a_0)$
yields $\lambda_0=\mu\xi/u^2$, so that
\begin{equation}
{\cal F}=\sqrt{\frac{\mu\xi}{2\pi\hbar u^2}}
\int_{-\infty}^{\infty}da_0\,\exp\left(-\frac{\cV_3(a_0/u)}{\hbar}\right).
\end{equation}
Changing the variable of integration to $z=a_0/u$
eliminates the unknown parameter $u$ from the problem,
leaving us with an expression for ${\cal F}$ which
depends only on the calculable parameters $\xi$ and $\mu$:
\begin{equation}
{\cal F}=\sqrt{\frac{\mu\xi}{2\pi\hbar}}
\int_{-\infty}^{\infty}dz\,\exp\left(-\frac{\cV_3(z)}{\hbar}\right).
\end{equation}

The case in which $S$ has only one extremum can be
dealt with similarly. Now $\cA'(a_0)$
has one real root ($a_0=0$), corresponding to the
minimum $\xgm(\tau)$ of $S[x]$, and a pair of
complex conjugate roots, $w$ and $w^*$, corresponding
to the pair of complex conjugate trajectories $\xct(\tau)$ and
$\xct^*(\tau)$. Correspondingly, we have
$\cA(a_0)=S[\xgm]+\cV_1(a_0/|w|)$, where
\begin{equation}
\label{U1}
\cV_1(z)\equiv\chi\left[\frac{1}{2}\,z^2
-\frac{2}{3}\,(\cos\phi)\,z^3+\frac{1}{4}\,z^4\right],
\end{equation}
with $\chi\equiv\alpha|w|^4$ and $\phi\equiv{\rm arg}(w)$.
Identifying $\cA(w)$ with $S[\xct]$ yields
\begin{equation}
S[\xct]-S[\xgm]=\frac{\chi}{12}
\left(2e^{2i\phi}-e^{4i\phi}\right),
\end{equation}
from which we can obtain
$\chi$ and $\phi$. Finally, identifying $\cV_1(a_0/|w|)$ with
$\cV(a_0)$ yields
$\lambda_0=\chi/|w|^2$, which leads to
\begin{equation}
{\cal F}=\sqrt{\frac{\chi}{2\pi\hbar|w|^2}}\,
\int_{-\infty}^{\infty}da_0\,\exp\left(-\frac{\cV_1(a_0/|w|)}{\hbar}\right)
=\sqrt{\frac{\chi}{2\pi\hbar}}\,
\int_{-\infty}^{\infty}dz\,\exp\left(-\frac{\cV_1(z)}{\hbar}\right).
\end{equation}

In the jargon of Ref.\ \cite{CAAC1}, the solution we have called
$\xlm(\tau)$ is a {\it one-saddle}: the operator of quadratic
fluctuations around it, $\hat{F}[\xlm]$, has only one negative
eigenvalue. At the second catastrophe, the one-saddle becomes
a {\it two-saddle} (i.e., it becomes unstable in another direction
in the functional space), and two one-saddles appear.
Having lower action than the two-saddle, the one-saddles have a larger
weight in the partition function. Hence, when the second caustic
is crossed the role of ``lowest saddle-point'' in Eqs.\ (\ref{xi})
and (\ref{lambda}) is transferred to one
of the newly born one-saddles, namely the one with the lowest action.
The transition is smooth as all three
trajectories coalesce and thus are identical to each other at the
caustic. We also notice that, in spite of the infinite number of
catastrophes, such a change of roles occurs
only once, namely at the second catastrophe, since
the minima of $S$ and the one-saddles
do not take part in the subsequent catastrophes. Indeed,
as shown in Ref.\ \cite{CAAC1} (see also Sec.\ \ref{infinity}), 
only $(n-1)$-saddles
and $n$-saddles take part in the $n$-th catastrophe.


\section{Application: the quartic double-well potential}
\label{application}

\subsection{Preliminaries}

Let us consider the quartic double-well potential, given by
\begin{equation}
V(x)=\frac{\lambda}{4}\,(x^2-a^2)^2
\qquad(\lambda>0).
\end{equation}
In order to simplify notation, it is convenient to
replace $x$ and $\tau$
by $q\equiv x/a$ and $\theta\equiv\omega\tau$,
respectively, with $\omega\equiv(\lambda a^2/m)^{1/2}$.
In the new variables, the equation of motion reads
$\ddot{q}=U'(q)$, where $U(q)\equiv\frac{1}{4}\,(q^2-1)^2$.
Its first integral is
\begin{equation}
\label{1stInt}
\frac{1}{2}\,\dot{q}^2=U(q)-U(\qt),
\end{equation}
where $\qt$ denotes the turning point (i.e., the point where
$\dot{q}=0$). This can be further integrated
to give us the relation between $\qt$ and the initial
position $q_0$ for a given ``time of flight''
$\Theta\equiv\beta\hbar\omega$. (As shown in Ref.\ \cite{CAAC2},
that relation is all we need in order to compute the
``usual'' semiclassical approximation to $\rho(x,x)$,
Eq.\ (\ref{sc}). The same is true for the computation
of the improved approximation, Eq.\ (\ref{exp2}), using the
procedure outlined in Sec.\ \ref{alternative}.)
Assuming for the moment
that $0\le q_0\le \qt\le 1$, we can write
\begin{equation}
\label{Theta}
\frac{\Theta}{2}=\int_{q_0}^{\qt}
\frac{dq}{\sqrt{2[U(q)-U(\qt)]}}.
\end{equation}
Inserting the explicit form of $U(q)$ and changing
the integration variable to $z=q/\qt$,
Eq.\ (\ref{Theta}) becomes
\begin{equation}
\label{u}
u=\int_{q_0/\qt}^1\frac{dz}{\sqrt{(1-z^2)(1-k^2z^2)}}\,,
\end{equation}
where
\begin{equation}
\label{uk}
u\equiv\frac{\Theta}{2}\sqrt{1-\frac{1}{2}\,\qt^2}\,,
\qquad k^2\equiv\frac{\qt^2}{2-\qt^2}\,.
\end{equation}
Performing the integration (formula 130.13 of Ref.\ \cite{Byrd})
and solving for $q_0$ finally yields
\begin{equation}
\label{q0qt}
q_0=\qt\,\cd(u,k),
\end{equation}
where $\cd$ is one of the Jacobian elliptic functions.

The action can be written as
$S[x]=(\hbar/g)\,I[q]$, where
$g\equiv\hbar\lambda/m^2\omega^3$ and
\begin{equation}
I[q]=\int_0^{\Theta}d\theta\left[\frac{1}{2}\,\dot{q}^2
+U(q)\right].
\end{equation}
Using Eq.\ (\ref{1stInt}), we may rewrite $I[\qc]$ as
\begin{equation}
\label{I}
I[\qc]=\Theta\,U(\qt)+2\int_{q_0}^{\qt}dq\,
\sqrt{2\,[U(q)-U(\qt)]}.
\end{equation}
The integration can be done with the help of formula 219.11 of
Ref.\ \cite{Byrd}. After a few algebraic manipulations one
arrives at
\begin{eqnarray}
\label{I[qc]}
I[\qc]&=&\Theta\,U(\qt)-\frac{1}{3}\,
\sqrt{2\,q_0^2\,(\qt^2-q_0^2)(2-\qt^2-q_0^2)}
\nonumber \\
& &-\frac{2}{3}\,\sqrt{2(2-\qt^2)}\left\{
(1-\qt^2)\,[\K(k)-\F(\varphi,k)]
-\E(k)+\E(\varphi,k)\right\},
\end{eqnarray}
where K, F and E are elliptic integrals \cite{GR,Byrd}
and $\varphi=\arcsin(q_0/\qt)$.

Equations (\ref{q0qt}) and
(\ref{I[qc]}) have been derived
under the assumption that $q_0$ and $\qt$ are real
and satisfy $0\le q_0\le \qt\le 1$.
However, since the elliptic functions and integrals are
meromorphic functions of their arguments,
we can now abandon that assumption and treat
$q_0$ and $\qt$ as complex variables.
(Note, however, that
$I[\qc]$ is a multivalued function of $\qt$ and so
one must be a bit careful when computing it.
For instance, the first square
root in Eq.\ (\ref{I[qc]}) acquires a minus sign
if $-1<\qt<-q_0<0$.)

Finally, the determinant of the fluctuation operator is given
by \cite{CAAC2}
\begin{equation}
\label{Delta}
\Delta=4\pi g\,{\rm sgn}(q_0-\qt)\,
\frac{\sqrt{2\,[U(q_0)-U(\qt)]}}
{U'(\qt)}\left(\frac{\partial q_0}{\partial \qt}
\right)_{\Theta}.
\end{equation}

We now have all the ingredients to compute the
semiclassical approximation to $\rho(x,x)$ --- both
the ``usual'' and the ``improved'' one.
Indeed, both the action and
the determinant of fluctuations can be
expressed solely in terms of $\qt$. Therefore,
as anticipated, this is the only information
we need from the classical trajectories.


\subsection{Singularities and their removal}
\label{sing}

\begin{figure}[htp]
\centerline{\hbox{\epsfig{figure=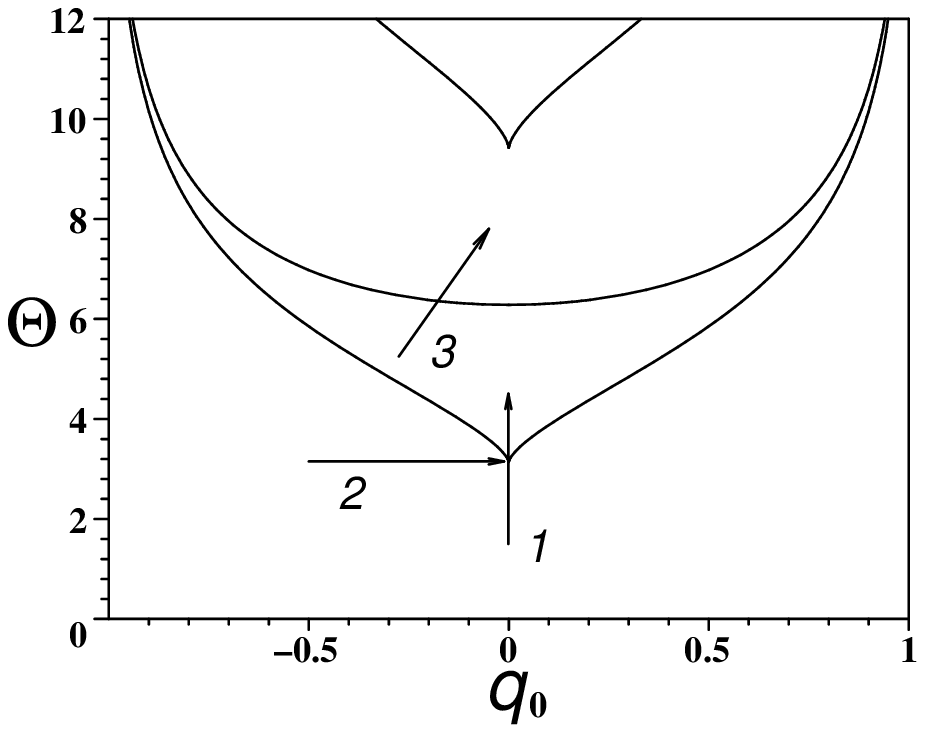,width=10cm}}}
\vspace{0.5cm}
\caption{The lower curve in this figure depicts the
caustic for the quartic double-well potential.
Below it the action has only one minimum; above it, the action
has two minima. The cusp is located at the point
$(q_0,\Theta)=(0,\pi)$. A second catastrophe occurs at
the curve in the middle: below it (but above
the caustic) the action has a one-saddle (in addition to the
minima); as the curve is crossed, the one-saddle splits into
a two-saddle and a pair of one-saddles, the latter corresponding
to a pair of periodic trajectories. The minimum of the middle curve is 
located at $(q_0,\Theta)=(0,2\pi)$. Upon crossing the upper curve,
whose minimum is located at $(q_0,\Theta)=(0,3\pi)$,
the number of classical trajectories increases by two: the two-saddle
splits into a three-saddle and a pair of two-saddles.
Numbered arrows
correspond to the first three subsections of Sec.\ \ref{sing}.
(This figure corrects Fig.\ 5 of Ref.\ \protect\cite{CAAC1}.)}
\label{caustic.ps}
\end{figure}

As we have already said, the ``usual'' semiclassical approximation
to $\rho(q_0,q_0)$ diverges at a caustic because of the
vanishing of the determinant of fluctuations $\Delta$
around the minimum of $S$ which appears or disappears there.
According to Eq.\ (\ref{Delta}), there are two ways $\Delta$
may vanish: (i) when $(\partial q_0/\partial \qt)_{\Theta}=0$;
(ii) when $U(q_0)=U(\qt)$.
A qualitative analysis of the equation of motion
shows that, at the boundary between the $N=1$ and the $N=2$
regions in the $(q_0,\Theta)$-plane, $\Delta$ vanishes
according to the first alternative \cite{CAAC1}. Solving the
equation $(\partial q_0/\partial \qt)_{\Theta}=0$ for
$\qt$ and inserting the result
$\tqt(\Theta)$ into Eq.\ (\ref{q0qt}), one obtains the lower
curve depicted in Fig.\ \ref{caustic.ps} --- the caustic.

In what follows we shall examine the behavior of
the ``usual'' semiclassical approximation across the
caustic, and compare it with the improved approximation.
(All numerical calculations were performed using 
\textsc{maple}.)


\subsubsection{$q_0=0$, $\Theta\approx\pi$}

When $q_0=0$ and $\Theta<\pi$, the only real
solution to Eq.\ (\ref{q0qt}) is $\qt=0$.
It then follows from Eq.\ (\ref{I}) that $I[\qc]=\Theta/4$.
In order to compute $\Delta$ we need $q_0(\qt,\Theta)$
for small $\qt$. Using Eqs.\ (\ref{uk}) and (\ref{q0qt}) we find
$q_0\approx \qt\,\cd(\Theta/2,0)=\qt\,\cos(\Theta/2)$;
Eq.\ (\ref{Delta}) then yields
$\Delta=2\pi g\,\sin\Theta$ in the limit $\qt\to 0$.
Therefore, the ``usual'' semiclassical approximation
to $\rho(0,0)$ gives
\begin{equation}
\rho_{\Sc}(0,0)\approx(2\pi g\,\sin\Theta)^{-1/2}\,
\exp\left(-\frac{\Theta}{4g}\right)
\qquad(\Theta<\pi).
\end{equation}
It diverges like $(\pi-\Theta)^{-1/2}$ as $\Theta\to\pi^-$.

\begin{figure}[htp]
\centerline{\hbox{\epsfig{figure=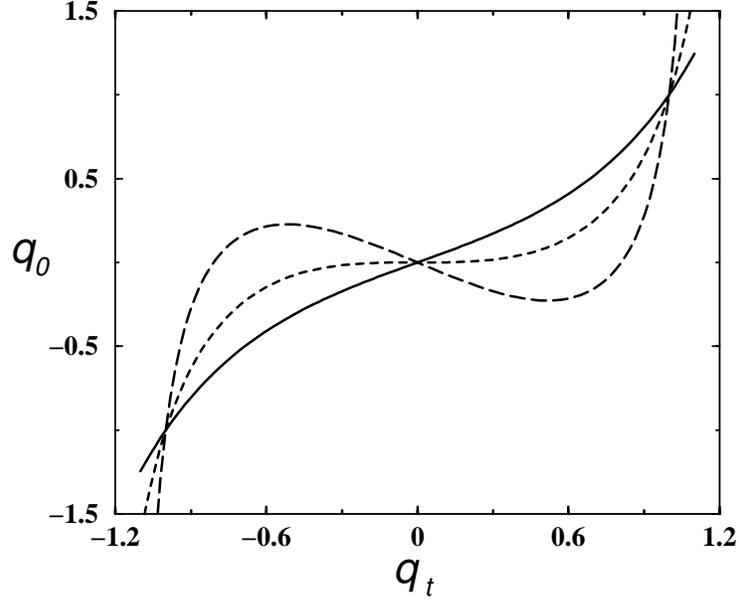,width=10cm}}}
\vspace{0.5cm}
\caption{$q_0(\qt,\Theta)$ [Eq.\ (\ref{q0qt})]
for $\Theta=2.0$ (solid line), $\Theta=\pi$ (short-dashed line),
and $\Theta=4.5$ (long-dashed line). For $\Theta<\pi$ this
function is one-to-one. For $\Theta>\pi$ and
$|q_0|$ sufficiently small there are three (or more) real 
values of $\qt$ corresponding to a given $q_0$.}
\label{q0qt.ps}
\end{figure}

\begin{figure}[htp]
\centerline{\hbox{\epsfig{figure=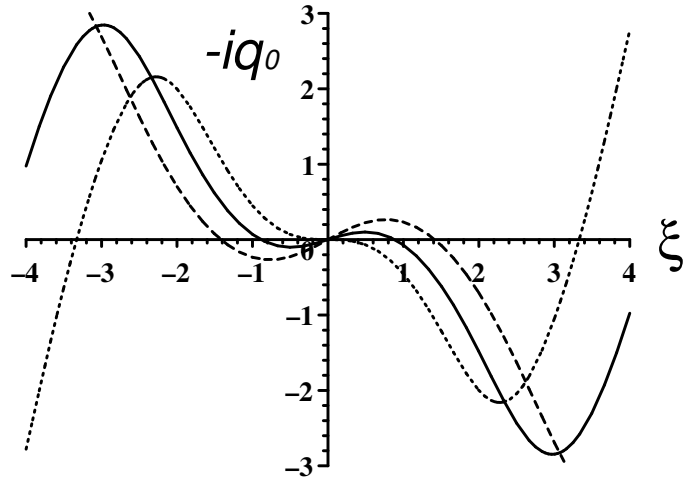,width=10cm}}}
\vspace{0.5cm}
\caption{$-iq_0(\xi,\Theta)$ [Eq.\ (\ref{q0xi})]
for $\Theta=2.0$ (solid line), $\Theta=2.5$ (short-dashed line),
and $\Theta=\pi$ (long-dashed line).}
\label{q0xi.ps}
\end{figure}

While for $\Theta<\pi$ there is only one real solution
to the equation $q_0(\qt,\Theta)=0$, for $\Theta>\pi$
there are three: $\qt=0$, corresponding to the trajectory
$\qc(\theta)\equiv 0$ (which is now a 1-saddle), plus a
pair of solutions located symmetrically with respect
to the origin,
corresponding to a pair of degenerate minima of the
action (see Fig.\ \ref{q0qt.ps}). The latter can be traced back to
a pair of purely imaginary trajectories for
$\Theta<\pi$. Indeed, making $\qt=i\xi$ in 
Eqs.\ (\ref{uk})--(\ref{q0qt}) and using the identity 
$\cd(u,ik)=\cn(u\,\sqrt{1+k^2},k/\sqrt{1+k^2})$ \cite{Byrd},
we obtain
\begin{equation}
\label{q0xi}
q_0=i\xi\,\cn\left(\frac{\Theta}{2}\,\sqrt{1+\xi^2},
\frac{\xi}{\sqrt{2(1+\xi^2)}}\right).
\end{equation}
The r.h.s.\ of the above equation
has an infinite number of zeros besides the one at
$\xi=0$ (see Fig.\ \ref{q0xi.ps}). As $\Theta$ approaches
$\pi$ from below, the zeros approach the origin and
two of them eventually coalesce there when
$\Theta=\pi$, reappearing as a pair of real solutions
to the equation $q_0(\qt,\Theta)=0$ for $\Theta>\pi$.

\begin{figure}[htp]
\centerline{\hbox{\epsfig{figure=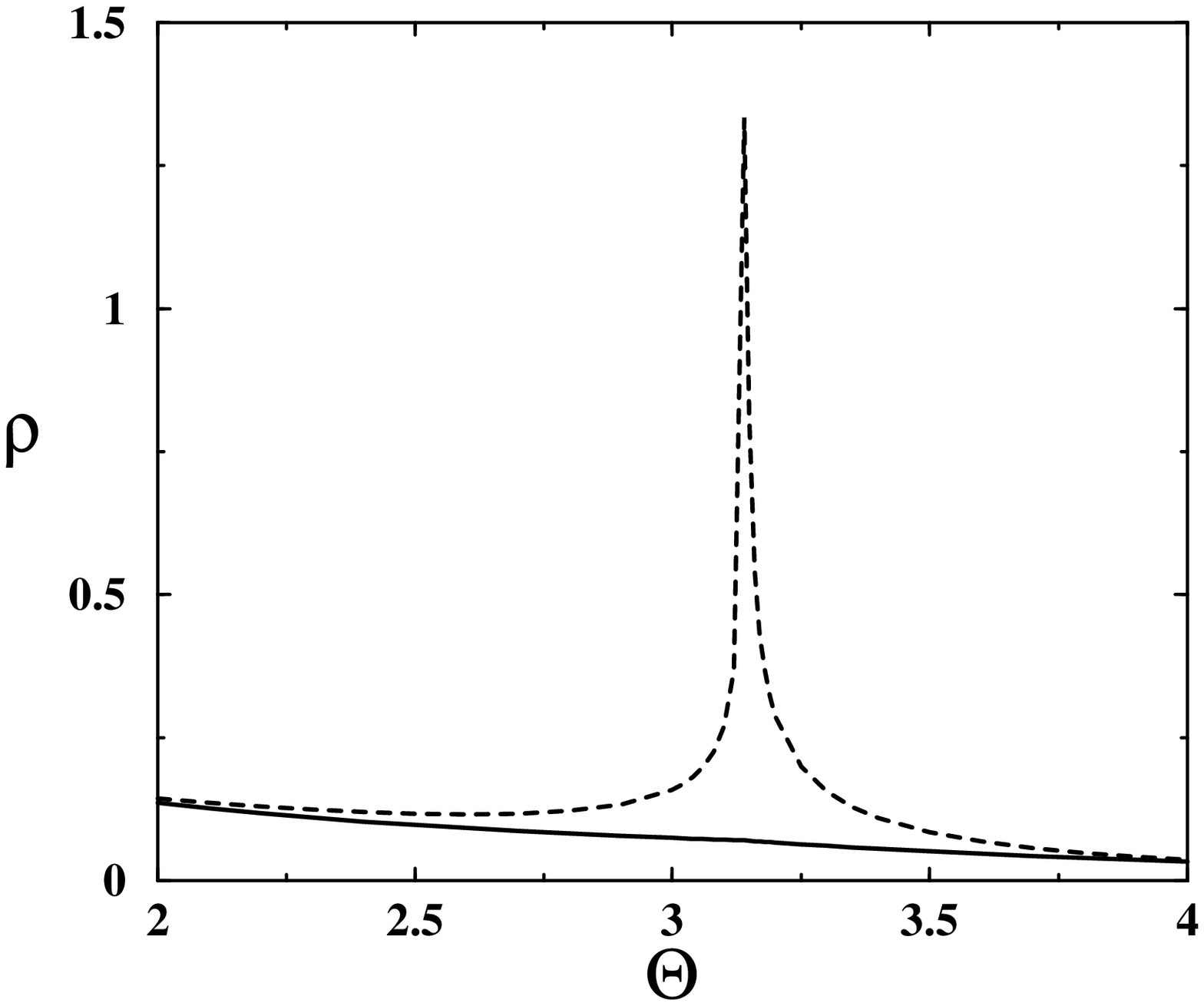,width=10cm}}}
\vspace{0.5cm}
\caption{$\rho(0,0)$ {\it vs}.\ $\Theta$ for $g=0.3$. Usual (dashed line) and
improved (solid line) semiclassical approximation.}
\label{rho_0.ps}
\end{figure}

In Fig.\ \ref{rho_0.ps} we show both the usual
[Eq.\ (\ref{sc})] and the improved [Eq.\ (\ref{exp2})]
semiclassical approximation to
$\rho(0,0)$ for $\Theta\approx\pi$. 


\subsubsection{$\Theta=\pi$, $q_0\approx 0$}

When $\Theta=\pi$, the approximation $q_0\approx \qt\,\cos(\Theta/2)$
is not enough for our purposes. Going to the next nontrivial order in the
Taylor expansion of $q_0(\qt,\pi)$ one obtains $q_0\sim \qt^3$
as $\qt\to 0$. It then follows from Eqs.\ (\ref{sc}), (\ref{I[qc]})
and (\ref{Delta}) that the usual semiclassical 
approximation to $\rho(q_0,q_0)$ behaves, for $\Theta=\pi$, as
\begin{equation}
\label{scpi}
\rho_{\Sc}(q_0,q_0)\stackrel{q_0\to 0}{\sim}g^{-1/2}\,
|q_0|^{-1/3}\,\exp\left(-\frac{\pi}{4g}\right).
\end{equation}
Two aspects of this result are worth of mention:
(i) the singularity at $q_0=0$ is integrable, hence the
semiclassical partition function is well defined; (ii) because
of the exponential factor, if $g\ll 1$ one has to be very
close to the origin to ``notice'' the singularity:
for $\rho_{\Sc}(q_0,q_0)$
to be of order unity or greater, $q_0$ must satisfy
$|q_0|\lesssim g^{-3/2}\,\exp(-3\pi/4g)$.

\begin{figure}[htp]
\centerline{\hbox{\epsfig{figure=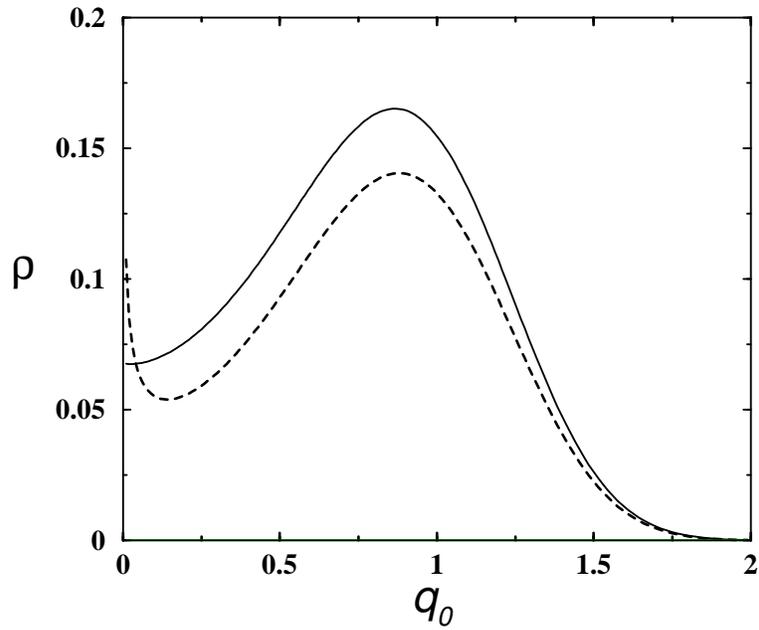,width=10cm}}}
\vspace{0.5cm}
\caption{$\rho(q_0,q_0)$ {\it vs}.\ $q_0$ for $\Theta=\pi$.
Usual (dashed line) and improved (solid line)
semiclassical approximation.}
\label{Pi.ps}
\end{figure}

Fig.\ \ref{Pi.ps} shows both the usual and the improved semiclassical
approximation to $\rho(q_0,q_0)$ for $\Theta=\pi$.
In order to make visible the singular behavior
of the former, we have taken $g=0.3$.
One can notice that far from the caustic (i.e., for 
$q_0\gtrsim 0.2$) the two curves are similar, 
but differ by approximately $10\%$.
This is due to the relatively large value of $g$.
The difference becomes smaller as $g$ is made smaller,
and vanishes in the classical limit $g\to 0$. 


\subsubsection{$\Theta>\pi$}

Expanding $q_0(\qt,\Theta)$ around $\tqt(\Theta)$,
we obtain $q_0-\tilde{q}_0\sim(\qt-\tqt)^2$,
so that $\partial q_0/{\partial \qt}\sim (\qt-\tqt)$
near the caustic. The other terms
in Eq.\ (\ref{Delta}) remain finite on it, so that
we finally obtain
\begin{equation}
\Delta^{-1/2}\sim |\qt-\tqt|^{-1/2}
\sim|q_0-\tilde{q}_0|^{-1/4}.
\end{equation}

\begin{figure}[htp]
\centerline{\hbox{\epsfig{figure=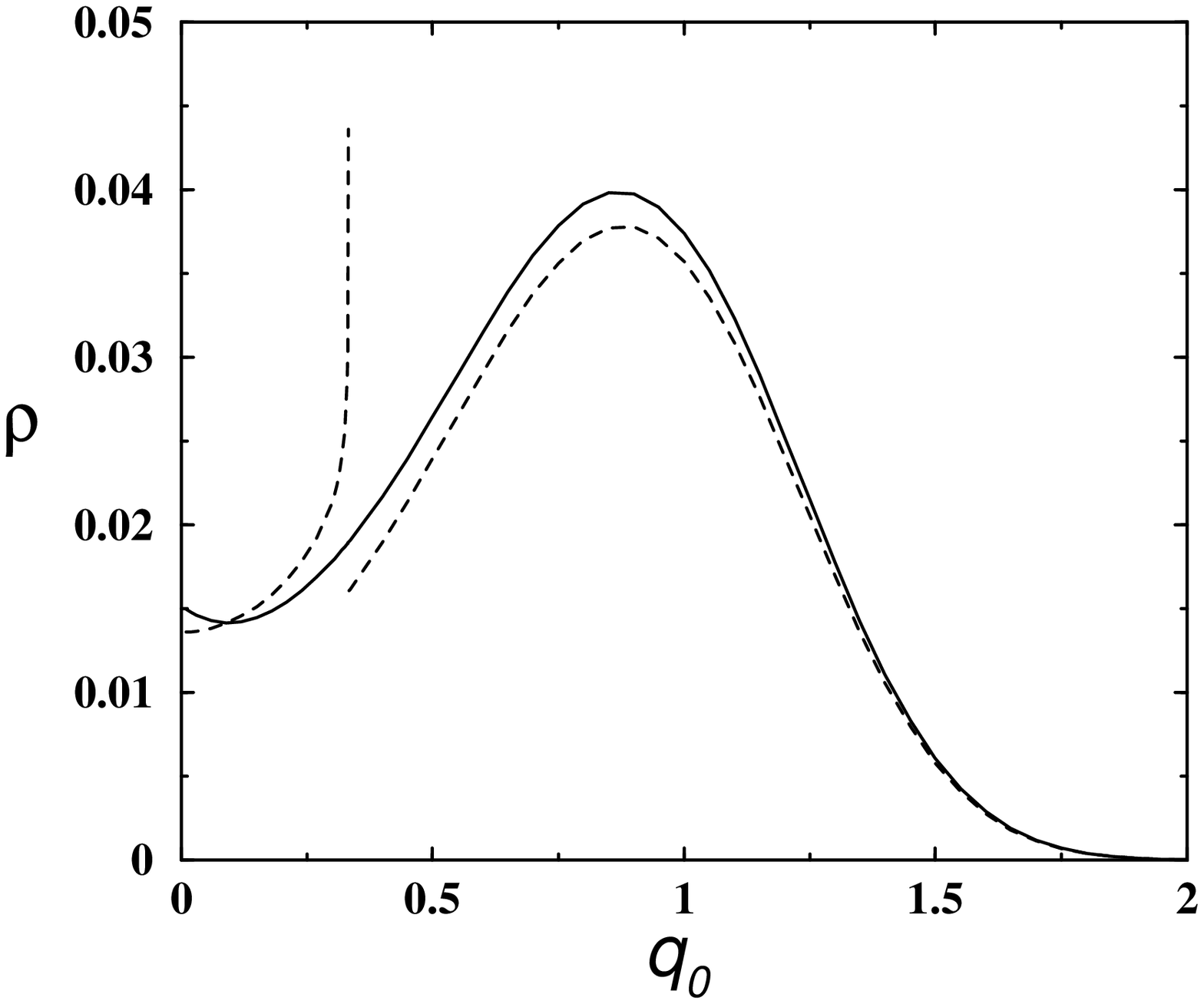,width=10cm}}}
\vspace{0.5cm}
\caption{$\rho(q_0,q_0)$ {\it vs}.\ $q_0$ for $\Theta=5.0$ and $g=0.3$.
Usual (dashed line) and improved (solid line) semiclassical
approximation.}
\label{Theta5.ps}
\end{figure}

\begin{figure}[htp]
\centerline{\hbox{\epsfig{figure=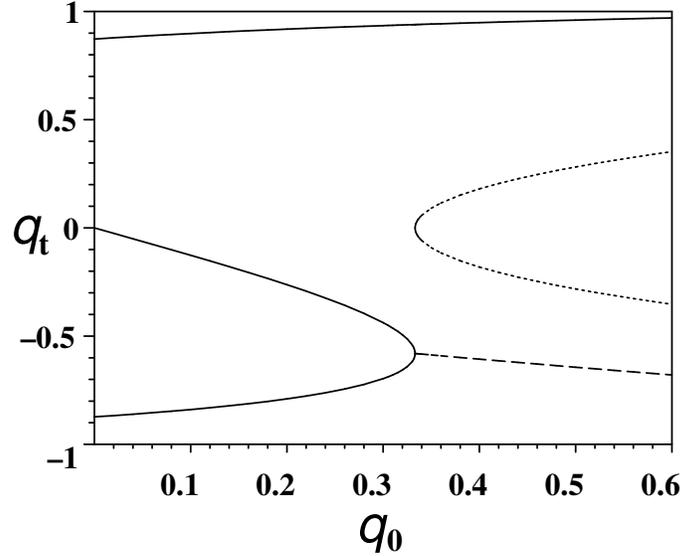,width=10cm}}}
\vspace{0.5cm}
\caption{$\qt$ {\it vs}.\ $q_0$ for $\Theta=5.0$.
If $|q_0|<q_*=0.3332\ldots$, there are three real solutions
to Eq.\ (\ref{q0qt}) (solid lines); the upper curve
corresponds to the global minimum, the lower one to the
local minimum, and the one in the middle to a one-saddle.
At the caustic the last two coalesce,
and reappear at the other side of the caustic (i.e., $|q_0|>q_*$) as a
pair of complex conjugate solutions. Their real and imaginary
parts are represented by the long-dashed and short-dashed lines,
respectively.}
\label{complex.ps}
\end{figure}

Fig.\ \ref{Theta5.ps} depicts both the usual and the
improved semiclassical approximation to
$\rho(q_0,q_0)$ for $\Theta=5.0$.
Again we had to use a relatively
large value of $g$ in order to magnify the
``critical'' region where the usual semiclassical
result diverges. Note that the divergence occurs only
at the two minima side of the caustic (see Fig.\ \ref{complex.ps}), 
as it is
associated with the coalescence of the local minimum
with a saddle-point of the action; the contribution
of the global minimum remains finite at the caustic.


\subsubsection{$\Theta\geq 2\pi$}
\label{2Pi}

As discussed in Ref.\ \cite{CAAC1}, another catastrophe is present
if $\Theta\ge 2\pi$. This time, $\Delta$ vanishes when the classical
trajectory is such that $U(q_0)=U(\qt)$, or, since the potential
is symmetric about the origin, $\qt=-q_0$. This catastrophe is
associated with the appearance of {\em periodic} classical
trajectories, and the condition $\qt=-q_0$ determines the
amplitude $A(\Theta)$ of these trajectories [$A(\Theta)$ is
the positive solution to equation $q_0(\qt,\Theta)=-\qt$,
where $q_0(\qt,\Theta)$ is defined by Eqs.\ (\ref{uk}) and
(\ref{q0qt})]. It is not difficult
to see why a catastrophe occurs when that condition is satisfied:
if $|q_0|<A(\Theta)$, there are two periodic trajectories
satisfying $\qc(0)=q_0$, related by time reversal, i.e.,
$\qc^{(2)}(\theta)=\qc^{(1)}(\Theta-\theta)$. If, on the other
hand, $|q_0|>A(\Theta)$, no such trajectories exist.
$|q_0|=A(\Theta)$ thus marks the boundary between regions with
zero and regions with two periodic trajectories. This boundary
is depicted in Fig.\ \ref{caustic.ps} (upper curve).

As discussed at the end of Sec.\ \ref{alternative}, 
the procedure for dealing with the caustic outlined in that section
is not affected by the appearence of a new catastrophe.
What changes as the second catastrophe is crossed 
is the identity of the ``lower saddle-point'': for
$|q_0|>A(\Theta)$, it is to be found among the solutions of 
Eq.\ (\ref{q0qt}); for $|q_0|<A(\Theta)$, it is given by
any one of the two periodic trajectories satisfying 
$\qc(0)=q_0$ (since they have the same action).
As a matter of fact, since all periodic trajectories with the
same amplitude (and the same period) have the same action, 
we may pick the one
that satisfies the condition $\qt=-q_0$, for then we
can use Eq.\ (\ref{I[qc]}) to compute its action.


\subsubsection{$\Theta\to\infty$}
\label{infinity}

In the high-temperature limit, $\Theta\to 0$, thermal wavelengths
are very small, and the classical limit sets in, as quantum
fluctuations are suppressed. That is the regime where
our improved semiclassical approximation should work best, as
illustrated in Ref.\ \cite{CAAC2}, since it incorporates quantum
fluctuations in a controlled manner as we lower the
temperature, and profits from the simplification of having to deal
with only one or two minima, as already emphasized. Nevertheless, even in
the opposite limit, $\Theta\to\infty$, our improved semiclassical
method can be used to reproduce zero temperature results. The secret
is to recognize that the various saddle-points which were discarded
for finite $\Theta$ {\em do} play a role in such a limit; in fact,
taking them into account is equivalent to using a dilute-gas
approximation, as we will qualitatively argue. First, however,
let us review how the saddle-points emerge.

For a fixed $q_0$ in the interval $(-1,+1)$, new saddle-points
will appear as we increase
$\Theta$, following a pattern outlined in Ref.\ \cite{CAAC1}.
As a result, the strip of the $(q_0,\Theta)$-plane defined by
$-1<q_0<+1$ and $\Theta\ge 0$ may be divided into regions
wherein each $(q_0,\Theta)$ point gives rise to $2n+1$ solutions,
$n=0,1,2,\ldots$, as shown in Fig.\ \ref{caustic.ps}. The regions
are separated by caustics where instabilities develop: for $n$
even, as we cross the caustic between the regions with $2n+1$
and $2n+3$ solutions, a $n$-saddle, i.e., a solution with $n$
negative eigenvalues, and a $(n+1)$-saddle appear; for odd
values of $n$, the $n$-saddle in the region with $2n+1$ solutions
becomes unstable --- it is replaced, in the region with $2n+3$
solutions, by two $n$-saddles which are periodic, and by a
$(n+1)$-saddle. Thus, for a given $\bar n$, we end up with a
$(\bar n+1)$-saddle, $(\bar n -1)$ pairs of $n$-saddles with
$n\le \bar n$, and a pair of minima (two $0$-saddles).

As $\Theta\to\infty$, the two minima will correspond to
solutions $\qc(\theta)$ that spend most of their euclidean
time $\theta$ near either $q_0=-1$ or $q_0=+1$. They both
have a single turning point, and only the local minimum will
cross the origin $q=0$: first, at a small value of $\theta$;
and upon returning, at a large value $\theta\sim\Theta$.
The $1$-saddles, which are periodic, have two turning points,
will also cross the origin twice, but one of the crossings
will occur at a value of $\theta$ near $\Theta/2$. Generalizing
this qualitative analysis, we may conclude that $n$-saddles
will have $(n+1)$ turning points and $n$ inner (away from
$\theta=0$ and $\theta=\Theta$) crossings of the origin.
Each of those inner crossings is equivalent to a kink or
an antikink, so that a generic $n$-saddle will not differ
much from a solution built out of a superposition of kinks
and antikinks in that limit. Typically, the euclidean time
width of such kinks should be much smaller than their
separations, as the solutions tend to spend most of their
time near $q_0=\pm 1$.

Varying the euclidean times where those inner crossings occur
should not alter significantly the action of the $n$-saddle
in the $\Theta\to\infty$ limit, an indication of the existence
of a flat direction in functional space corresponding to that
variation. As flat directions are associated to near-zero
eigenvalues, we claim that the negative eigenvalues which
characterize the $n$-saddle will approach zero from below
as $\Theta\to\infty$, and that the euclidean times where
the crossings occur should be treated as collective coordinates \cite{RR},
just as the positions of kinks and antikinks in the dilute-gas
approximation. The contributions of the various kinks and
antikinks can be dealt with in the usual manner --- they add
up to an exponential, and reproduce the standard result for
the splitting between ground and first-excited state \cite{coleman}.
(See, however, Ref.\ \cite{Rossi} for a treatment of the
low-temperature limit within the functional formalism that does
not appeal to the dilute-gas approximation.)


\section{Conclusion}
\label{conclusion}

Semiclassical methods are a powerful nonperturbative tool, for
both equilibrium
and nonequilibrium systems. This article, together with Refs.
\cite{CAAC2,CAAC3,CAAC1}, represents a further step towards a
systematic semiclassical treatment of quantum statistical mechanics.

In the present work, we developed a simple procedure to derive
the lowest order semiclassical approximation for the case of
multiple-well potentials in equilibrium quantum statistical
mechanics. In order to adequately incorporate new extrema, we
kept fluctuations beyond the quadratic level along the
``unstable'' direction in functional space, and relied on our
knowledge of the type of catastrophe involved as we cross
a caustic to eliminate spurious singularities in the semiclassical
approximation,
obtaining sensible results for the density matrix elements for any
temperature.
This was exemplified by the analysis of the quartic
double-well potential.

Our results can possibly be extended to nonequilibrium systems,
such as those where a time-dependent
potential is coupled to a heat bath, in order to better understand
transient regimes. Although the physics of nonequilibrium quantum
statistical mechanics has been considered in detail in the context
of semiclassical calculations of the decay rates of metastable systems
\cite{Affleck,Chudnovsky,Weiss,Ankerhold,Weiper1}, a thorough analysis
of the various transient regimes,
and of the interplay of their corresponding
time scales, is still needed. Here, however, we will no longer profit
from the drastic reduction in the number of extrema that occurs in equilibrium
situations, as time evolution forces us to deal with saddle points and
maxima, as
well. The simplified methods presented in this paper will still be
useful to describe
the asymptotic imaginary time evolution corresponding to equilibrium,
but not the
real time evolution, which requires the traditional quantum mechanical
treament.

As for possible extensions to field theories, the methods developed
in \cite{CAAC2,CAAC3}
should be applicable to the evaluation of the effective potential
in the presence of non-trivial backgrounds (defects), as long as
they depend on only one coordinate. This can be of use in a wealth
of possible applications, and should help
in the study of phase transitions and critical phenomena where such
defects play a role. Cases such as the ones explored here and
in \cite{CAAC1}, which involve several extrema, still lack a
field theoretic treatment.


\acknowledgements

E.S.F. acknowledges Valerio Tognetti, Ruggero Vaia and
Alessandro Cuccoli for their kind hospitality in his visit to
Universit\`a di Firenze and for fruitful discussions.
The authors acknowledge support from CNPq (C.A.A.C., E.S.F.\
and S.E.J.), FUJB/UFRJ (C.A.A.C.), FAPESP and FAPERJ (R.M.C.).
E.S.F.\ and S.E.J.\ were partially supported by the U.S.\
Department of Energy under the contracts DE-AC02-98CH10886 and
DE-FG02-91ER40688 - Task A, respectively.



\newpage

\end{document}